# Magnetic tunnel junctions with ferroelectric barriers: Prediction of four resistance states from first-principles

Julian P. Velev,[1,2 *†] Chun-Gang Duan,[3†] J. D. Burton,[1] Alexander Smogunov,[4,5] Manish K. Niranjan,[1] Erio Tosatti,[4,5] S. S. Jaswal,[1] and Evgeny Y. Tsymbal [1*]

[1] *Department of Physics and Nebraska Center for Materials and Nanoscience, University of Nebraska, Lincoln, NE 68588-0111, USA*
[2] *Department of Physics and Institute for Functional Nanomaterials, University of Puerto Rico, San Juan, PR 00931-3344, USA*
[3] *Key Laboratory of Polarized Materials and Devices, East China Normal University, Shanghai 200062, China*
[4] *International Centre for Theoretical Physics (ICTP), Strada Costiera 11, 34014 Trieste, Italy*
[5] *International School for Advanced Studies (SISSA) and CNR/DEMOCRITOS National Simulation Center, Via Beirut 2-4, 34014 Trieste, Italy*

* Corresponding authors: J.P.V. (e-mail: jvelev@gmail.com) or E.Y.T. (e-mail: tsymbal@unl.edu)
† Co-first authors

Magnetic tunnel junctions (MTJs), composed of two ferromagnetic electrodes separated by a thin insulating barrier layer, are currently used in spintronic devices, such as magnetic sensors and magnetic random access memories. Recently, driven by demonstrations of ferroelectricity at the nanoscale, thin-film ferroelectric barriers were proposed to extend the functionality of MTJs. Due to the sensitivity of conductance to the magnetization alignment of the electrodes (tunnelling magnetoresistance) and the polarization orientation in the ferroelectric barrier (tunnelling electroresistance), these *multiferroic tunnel junctions* (MFTJs) may serve as four-state resistance devices. Based on first-principles calculations we demonstrate four resistance states in $SrRuO_3/BaTiO_3/SrRuO_3$ MFTJs with asymmetric interfaces. We find that the resistance of such a MFTJ is significantly changed when the electric polarization of the barrier is reversed and/or when the magnetizations of the electrodes are switched from parallel to antiparallel. These results reveal the exciting prospects of MFTJs for application as multifunctional spintronic devices.

The field of spintronics has been successful in producing magnetoresistive devices for magnetic memory and sensor applications.[1] These employ giant magnetoresistance (GMR)[2,3] or tunnelling magnetoresistance (TMR)[4-6] phenomena that provide a sizable change of resistance in response to altering magnetic alignment of two ferromagnetic electrodes in a magnetic metallic multilayer or a magnetic tunnel junction (for reviews see refs. 7 and 8). The evolution beyond passive magnetoelectronic components is envisioned in the next generation of multifunctional spintronic materials and structures whose properties can be manipulated by several independent stimuli by affecting physical degrees of freedom set by the order parameters.[9-11] Multiferroic tunnel junctions (MFTJs), which exploit an active ferroelectric barrier in a tunnel junction with ferromagnetic electrodes, serve as examples of such systems.[12]

Until recently the idea to combine ferroelectricity and electron tunnelling in a single device seemed to be unfeasible because quantum tunnelling is only possible through barrier thickness less than a few nanometres, the scale at which ferroelectricity was thought not to exist. Recent experimental and theoretical findings demonstrate, however, that ferroelectricity persists down to atomic sizes.[13-15] In particular, it was discovered that in organic ferroelectrics ferroelectricity can be sustained in thin films of monolayer thickness.[16] In perovskite ferroelectric oxides ferroelectricity was observed down to the nanometre scale.[17-21] These experimental results are consistent with first-principles calculations that predict that the critical thickness for ferroelectricity in perovskite films can be as small as a few lattice parameters.[22-26]

The existence of ferroelectricity in nanometre-thick films makes it possible to use ferroelectrics as barriers in tunnel junctions.[12,27] In such *ferroelectric tunnel junctions* (FTJs) the ferroelectric polarization of the barrier can be reversed by an external electric field producing an additional degree of freedom that may be explored in novel electronic devices. The basic idea of a FTJ (called a polar switch) was formulated by Esaki *et al.*[28] However, experimental investigations of FTJs have started only recently, driven by the discovery of ferroelectricity in ultrathin films.[29] Theoretical studies of FTJs indicate that the reversal of the electric polarization of the ferroelectric barrier can produce a sizable change in resistance or *tunnelling electroresistance* (TER) effect.[30,31]

The functional properties of FTJs can be extended by replacing metal electrodes by ferromagnetic metals, making the junction *multiferroic*. In such a MFTJ, the spin-dependent tunnelling may be controlled through manipulation of the electric polarization of the ferroelectric barrier.[32,33] The coexistence of TER and TMR effects in a MFTJ makes it a *four-state resistive device*. An alternative route to MFTJs is to employ single-phase multiferroic materials, such as $La_{0.1}Bi_{0.9}MnO_3$, as a barrier.[34] However, because single-phase multiferroics are very rare in nature and just a few of them retain multiferroic properties at room temperature,[35] it is advantageous to make MFTJs from the combination of ferroelectric and ferromagnetic materials.

In this paper, we use first-principles calculations based on density functional theory to demonstrate the coexistence of TMR and TER effects in a single MFTJ. We find four resistance states of a MFTJ that are distinguished by the orientation of the electric polarization of the barrier (left or right) and the relative magnetization of the electrodes (parallel or antiparallel). We consider a $SrRuO_3/BaTiO_3/SrRuO_3$(001) MFTJ as a model system to explore this phenomenon. This choice is motivated by the fact that barium titanate, $BaTiO_3$, is the prototypical perovskite ferroelectric oxide. Thin $BaTiO_3$ films have been shown to retain ferroelectric properties down to one unit cell thickness.[20] Strontium ruthenate, $SrRuO_3$, is a ferromagnetic metal ($T_C$~160K) which has the same perovskite crystal structure as $BaTiO_3$ with lattice mismatch less than 2%. Recently, epitaxial $SrRuO_3/BaTiO_3/SrRuO_3$ (001) structures have been fabricated which exhibit sizable polarization of $BaTiO_3$ at thicknesses of a few nanometres,[36,37] indicating that such MFTJs are experimentally feasible.

The atomic and electronic structure of $SrRuO_3/BaTiO_3/SrRuO_3$ MFTJs is obtained by first-principles calculations based on density functional theory. To simulate large enough systems we use a plane-wave pseudopotential approach as implemented in the Quantum-ESPRESSO package (QE)[38] and the Vienna *Ab-Initio* Simulation Package (VASP).[39] Fig. 1 shows the atomic structure of the junction with six unit cells of $BaTiO_3$ placed between $SrRuO_3$ electrodes. The $BaTiO_3$ film has different interface terminations: BaO at the left and $TiO_2$ at the right interface. The presence of dissimilar interfaces makes the MFTJ asymmetric, a necessary requirement for observing the TER effect.[30] Such control over the interface terminations is experimentally feasible due to ionic oxides exhibiting unit cell by unit cell growth.[37] The in-plane lattice constant is fixed to be the experimental value of the bulk $BaTiO_3$ ($a$ = 3.991 Å), which is smaller than the theoretical lattice constant of $BaTiO_3$ (4.033 Å) but larger than that of $SrRuO_3$ (3.97 Å). This produces a tetragonal distortion of both $SrRuO_3$ ($c/a$ = 0.988) and $BaTiO_3$ ($c/a$ = 1.021), favouring electric polarization perpendicular to the plane of the layers.

First, we find equilibrium atomic structures of the MFTJ corresponding to the two polarization states of $BaTiO_3$ with polarization pointing to the right ($P_\rightarrow$) or to the left ($P_\leftarrow$).[40] The ferroelectric instability of bulk $BaTiO_3$ is characterized by uniform displacements of the cations (Ti and Ba) with respect to the O anions (soft mode). In bulk $BaTiO_3$ the two opposite polarization directions are equivalent resulting in the symmetric double potential well profile shown schematically by the dashed line in Fig. 2a.[41] For a thin film, the soft mode displacements are constrained by interface relaxations which depend on the structure of the interfaces.[24] The interface ionic and electronic relaxations produce interface dipoles that are present even in the paraelectric state of $BaTiO_3$ (black curves in Fig. 1). The dipoles are oriented towards the barrier at both interfaces, but they are very different in magnitude. The Ru displacement on the left is much larger than that of Ti on the right due to the larger ionic radius of the Ba and Ru atoms compared to Sr and Ti. This constrains the relaxation of the interface Ti atom in the $BaTiO_3$ and produces asymmetry between the $P_\rightarrow$ and $P_\leftarrow$ polarization states. In addition, there is an intrinsic electric field throughout the bulk of the barrier produced by the dissimilar electrostatic dipoles at the two interfaces. The combined effect of the geometrical constraints at the interfaces and the intrinsic electric field can be thought of as a total *effective* electric field that causes the two polarization states to have different energies, making the potential profile asymmetric,[42] as is indicated schematically by the solid line in Fig. 2a.

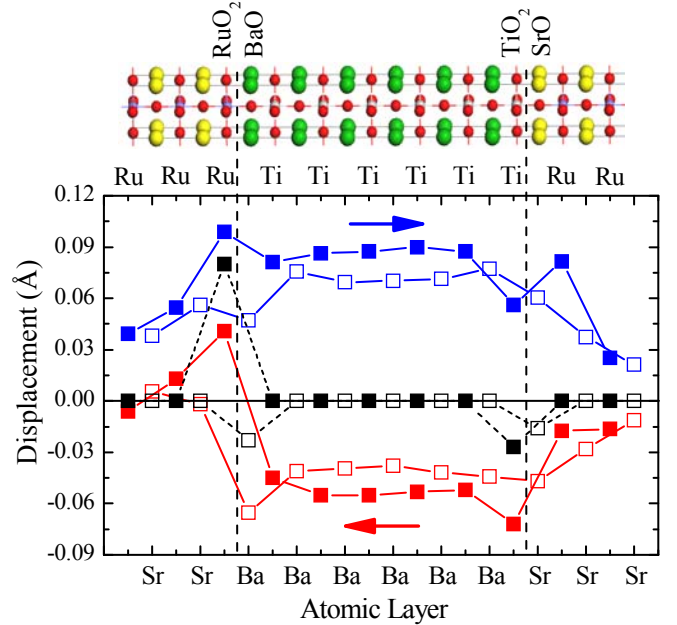

**Figure 1: Atomic structure of the $SrRuO_3/BaTiO_3/SrRuO_3$ MFTJ.** Displacements of the cations (Sr-Ba, Ru-Ti) are measured with respect to the O atoms in the same plane. Solid symbols denote displacements of Ru and Ti; open symbols denote displacements of Sr and Ba. Red/blue curves correspond to polarization of $BaTiO_3$ pointing to the left/right. Black curves show interface relaxations for paraelectric $BaTiO_3$. The interfaces are indicated by vertical dashed lines.

For $BaTiO_3$ thickness of six unit cells, we find that the energy difference between the $P_\leftarrow$ and $P_\rightarrow$ states is 30 meV. The $P_\rightarrow$ state is characterized by displacements of Ti ions with respect to O ions at the central $TiO_2$ layers of about 0.09 Å. These are comparable to the respective displacements in bulk $BaTiO_3$ (0.12 Å). The unfavourable effective field pointing to the right prevents the $P_\leftarrow$ state to develop until the thickness of $BaTiO_3$ reaches six unit cells. Even then the polarization displacements of 0.06 Å are substantially smaller than those for the $P_\rightarrow$ state. Thus, the different interface terminations produce asymmetric ferroelectric displacements which play an important role in the TER effect. The obtained structural properties and energetics of the $BaTiO_3/SrRuO_3$ heterostructure with asymmetric interfaces are consistent with previous calculations.[42]

Polarization charges at the interfaces are not completely screened by the $SrRuO_3$ electrodes. The incomplete screening gives rise to a depolarizing field pointing opposite to the polarization. Fig. 2b shows the electrostatic potential energy profile within the junction. The effect of the electrostatic potential is a gradual decrease in energy of the $BaTiO_3$ conduction band minimum (CBM) from about 0.4 to 0.7 eV with respect to the Fermi level in the direction of polarization. This is reflected in the local density of states on the interfacial Ti atoms, shown in Fig. 2c. The shift is in the opposite direction for the $P_\leftarrow$ and $P_\rightarrow$ states.

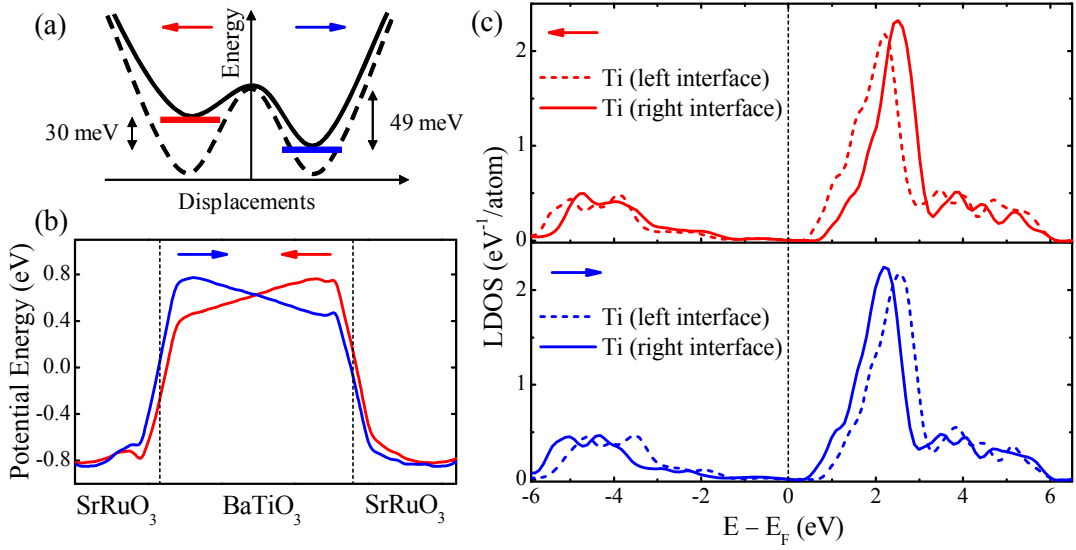

**Figure 2: Effects of ferroelectricity in the SrRuO₃/BaTiO₃/SrRuO₃ MFTJ.** (**a**) Schematic double-well potential for bulk BaTiO₃ (dashed line) and for the MFTJ (solid line). (**b**) Cell averaged electrostatic potential energy profile for polarization to the right $P_\rightarrow$(blue) and left $P_\leftarrow$(red) states. Interfaces are indicated by vertical dashed lines. (**c**) Local density of states (LDOS) on the left (dashed line) and right (solid line) interfacial Ti atoms for $P_\rightarrow$ (top panel) and $P_\leftarrow$ (bottom panel) states. The Fermi energy is denoted by the vertical line.

Next, the tunnelling conductance of the MFTJ is calculated using a general scattering formalism[43,44] implemented in QE. The structure of Fig. 1 is considered as the scattering region, ideally attached on both sides to semi-infinite SrRuO₃ leads.[45] The conductance per unit cell area is given by the Landauer-Büttiker formula

$$G = \frac{e^2}{h} \sum_{\sigma \mathbf{k}_\parallel} T_\sigma(\mathbf{k}_\parallel), \quad (1)$$

where $T_\sigma(\mathbf{k}_\parallel)$ is the transmission probability of the electron with spin $\sigma$ at the Fermi energy. The $\mathbf{k}_\parallel = (k_x, k_y)$ is the Bloch wave vector corresponding to the periodicity in the plane of the junction. Since the $P_\leftarrow$ and $P_\rightarrow$ states are non-equivalent with respect to inversion, four distinct conductance states for the MFTJ are then produced by two TMR and two TER conductance states. Fig. 3 shows schematically the four conductance levels indicating the possibility of switching between them by electric (**E**) and magnetic (**H**) fields.

The figures of merit for the MFTJ are the magnitudes of the tunnelling magnetoresistance and electroresistance effects. We define the TMR ratio as [46]

$$\text{TMR} = \frac{G_{\uparrow\uparrow} - G_{\uparrow\downarrow}}{G_{\uparrow\uparrow} + G_{\uparrow\downarrow}}, \quad (2)$$

where $G_{\uparrow\uparrow}$ is the conductance of the parallel and $G_{\uparrow\downarrow}$ is the conductance of the antiparallel magnetization configuration. The TER ratio is defined as

$$\text{TER} = \frac{G_\leftarrow - G_\rightarrow}{G_\leftarrow + G_\rightarrow}, \quad (3)$$

where $G_\rightarrow$ is the conductance for polarization pointing to the right and $G_\leftarrow$ the conductance for polarization pointing to the left. Both the calculated TMR and TER ratios are very large and are dependent on the other ferroic order (i.e. TMR depends on the polarization state and TER depends on the magnetization state). This is the signature of a true multifunctional device.

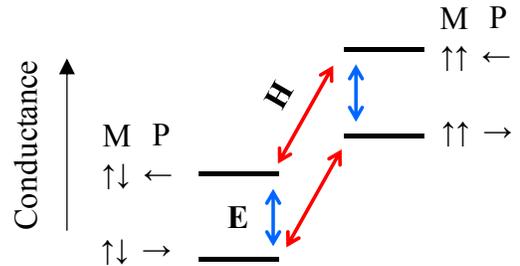

| G (10⁻⁷ e²/h) | ↑↑ | ↑↓ | ↑↑:↑↓ | TMR (%) |
|---|---|---|---|---|
| → | 3.76 | 0.83 | 4.6 | **64** |
| ← | 11.82 | 1.69 | 7.0 | **75** |
| ← : → | 3.1 | 2.1 | | |
| **TER (%)** | **52** | **35** | | |

**Figure 3: Conductance of the SrRuO₃/BaTiO₃/SrRuO₃ MFTJ.** The four conductance states are distinguished by polarization in the barrier pointing to the left (←) or right (→) and magnetization of the electrodes being parallel (↑↑) or antiparallel (↑↓). Conductance values are given per transverse area of the unit cell. TMR and TER ratios are defined according to eqs. (2) and (3), respectively. The diagram on the top shows schematically the four resistance states that can be controlled by electric (**E**) and magnetic (**H**) fields.

To gain insight into the large changes in the conductance as the various ferroic order parameters are switched, we analyze the $\mathbf{k}_\parallel$-resolved transmission, plotted in Fig. 4. When the magnetizations of the electrodes are aligned parallel, for either spin channel or polarization orientation the main contribution to the conductance comes from the area around the Brillouin zone centre at $\mathbf{k}_\parallel = \mathbf{0}$ ($\bar{\Gamma}$ point). In a generic cubic perovskite crystal, $ABO_3$, the $\Gamma$-point $d$ states of the transition metal B atom are split by the octahedral crystal field produced by the O cage into $t_{2g}$ states ($d_{zx}$, $d_{zy}$, $d_{xy}$) and $e_g$ states ($d_{z^2}$ and $d_{x^2-y^2}$), with the $t_{2g}$ levels lying lower in energy. Hybridization with the O $p$ states splits these $d$ bands into low lying bonding bands, and high lying anti-bonding bands. The anti-bonding bands either become the conduction bands of insulating perovskites (such as $BaTiO_3$), or form the Fermi surface of metallic perovskites (such as $SrRuO_3$). In the layered perovskites system we consider here, the symmetry is lowered from cubic to that of a square, and at the $\bar{\Gamma}$ point the bands can be categorized by their axial symmetry around the $z$-axis: $t_{2g}$ bands split to form a doubly degenerate band with $\Delta_5$ symmetry ($d_{zx}$, $d_{zy}$) and another band having $\Delta_2$ axial symmetry ($d_{xy}$). Similarly the $e_g$ bands split into one band with $\Delta_1$ symmetry ($d_{z^2}$) and one with $\Delta_{2'}$ symmetry ($d_{x^2-y^2}$).

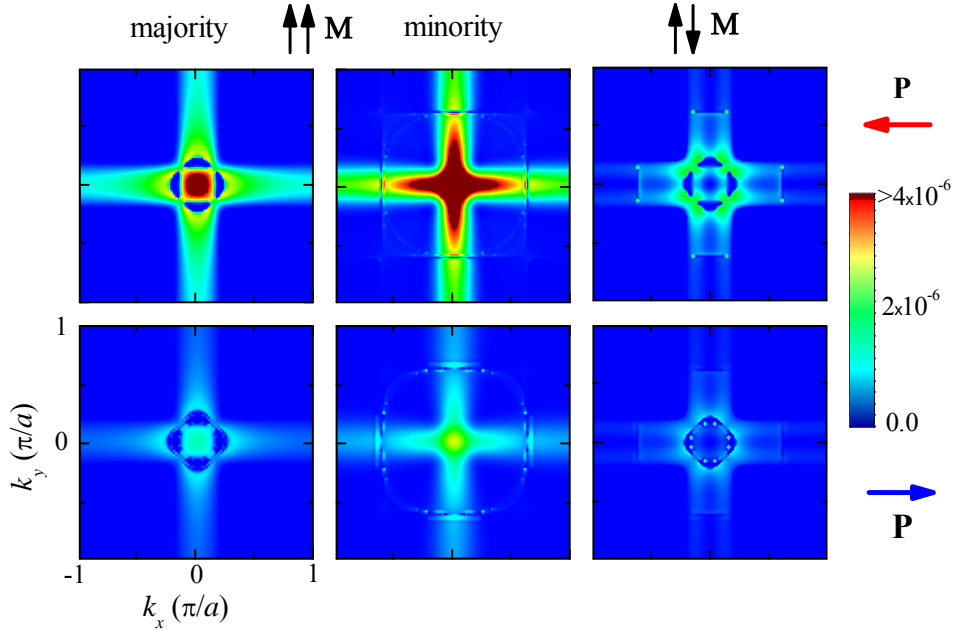

**Figure 4: Transmission in the 2D Brillouin zone of the $SrRuO_3/BaTiO_3/SrRuO_3$ MFTJ.** Top (bottom) panels show $\mathbf{k}_\parallel$-resolved transmission for polarizations of the $BaTiO_3$ barrier pointing to left (right). Here are the components of the in-plane Bloch wave vector $\mathbf{k}_\parallel$. Left and middle panels show majority- and minority-spin electrons, respectively, for parallel magnetization of the electrodes. Right panels show antiparallel magnetization of the electrodes.

Due to the exchange splitting in the ferromagnetic metal perovskite $SrRuO_3$, the majority $t_{2g}$ anti-bonding bands are almost fully occupied and the minority $e_g$ anti-bonding bands are completely empty. Conversely, the majority $e_g$ and the minority $t_{2g}$ anti-bonding bands are partially occupied, forming the Fermi surface. Along the [001] direction at $\bar{\Gamma}$ there is one Fermi sheet in the majority spin channel consisting of states with $\Delta_1$ symmetry and two degenerate Fermi sheets in the minority channel with $\Delta_5$ symmetry (see Fig. 5a). As a result, the conduction properties of the $SrRuO_3$ electrodes in our tunnel junctions are controlled by one propagating state in the majority- and two in the minority-spin channel at the Brillouin zone centre.

In $BaTiO_3$ the anti-bonding states are completely empty and the insulating gap reflects the splitting between the bonding and anti-bonding bands, with the $t_{2g}$ anti-bonding bands forming the lowest conduction bands. Within the band gap there exist evanescent states with wavefunctions that decay exponentially with a rate $\kappa$ determined by the complex band structure.[47] These complex bands can also be characterized by their symmetry: the three states with the lowest decay rates within the band gap consist of a $\Delta_5$ doublet and a $\Delta_1$ singlet. Importantly, these complex bands are highly sensitive to the magnitude of the ferroelectric displacements which, in turn, depend significantly on the polarization direction in the barrier (see Fig. 1). This is illustrated in Fig. 5b by plotting the complex bands of bulk $BaTiO_3$ for two different soft-mode magnitudes corresponding to the displacements at the centre of the barrier for the $\mathbf{P}_\leftarrow$ and $\mathbf{P}_\rightarrow$ states. It is seen that the main effect is an overall increase in the decay constants with larger soft-mode magnitude, consistent with the previous comparison between ferroelectric and paraelectric $BaTiO_3$.[31] This arises due to the reduced dispersion of the conduction bands states as the Ti atoms are displaced closer to one of the O atoms, yielding a corresponding increase in the band gap.

Owing to the preservation of wavefunction symmetry across the epitaxial SrRuO$_3$/BaTiO$_3$ interfaces, the $\bar{\Gamma}$ point majority-spin states at the Fermi level decay inside the barrier according to the $\Delta_1$ band of BaTiO$_3$, whereas the minority-spin states decay according to the $\Delta_5$ band, yielding a perfect correspondence between symmetry and spin. In the parallel magnetic configuration, both polarization directions conduct with the minority-spin conductance being approximately twice as large as the majority-spin conductance (see Fig. 4). For the antiparallel magnetic configuration, there is no conductance at the $\bar{\Gamma}$ point due to symmetry mismatch. Majority-spin $\Delta_1$ states of the left electrode cannot be transmitted to the minority-spin $\Delta_5$ states of the right electrode and vice versa. As a result the antiparallel conductance is much smaller than the parallel, yielding the large TMR values listed in Fig. 3.

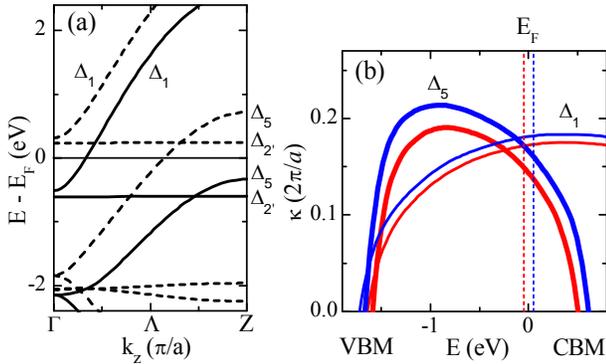

**Figure 5: Electronic band structure of SrRuO$_3$ and BaTiO$_3$.** (**a**) Spin-polarized bands along the [001] direction for SrRuO$_3$. Majority-spin (solid) and minority-spin (dashed) bands near the Fermi energy are labelled with their symmetry. (**b**) Decay constant for BaTiO$_3$ for soft-mode displacement magnitudes corresponding to the $\mathbf{P}_\leftarrow$ (red) and the $\mathbf{P}_\rightarrow$ (blue) states. Thick curves indicate the doubly degenerate $\Delta_5$ states, and the thin curves the $\Delta_1$ symmetry states. Vertical dashed lines show the average position of the Fermi energy in the tunnel junction with respect to the valence band maximum (VBM) and the conduction band minimum (CBM).

To elucidate the effect of ferroelectric polarization on the conductance we approximate the transmission probability for an electron within a given band and $\mathbf{k}_\parallel$ by [48]

$$T(\mathbf{k}_\parallel) \approx t_L(\mathbf{k}_\parallel)\exp\left[-2\kappa(\mathbf{k}_\parallel)d\right]t_R(\mathbf{k}_\parallel). \quad (4)$$

Here $t_L(\mathbf{k}_\parallel)$ and $t_R(\mathbf{k}_\parallel)$ are the interface transmission functions (ITF) characterizing the left and right interfaces. The exponential factor accounts for the decay of the electronic wavefunction through the barrier of thickness $d$ with a decay rate $\kappa(\mathbf{k}_\parallel)$. According to eq. (4), there are two possible factors contributing to the polarization dependence of the transmission: (i) change of the ITFs and/or (ii) change of the decay rate in the barrier.

Due to the electrostatic shift of the CBM, the barrier height at the interface decreases along the polarization direction. Therefore, when polarization is pointing to the right, $t_L$ increases and $t_R$ decreases; and the other way around when polarization is to the left. Thus the product $t_L t_R$ does not change significantly, therefore, it cannot account for the factor of 3 change in conductance.

Changes in the decay rate induced by polarization reversal can be estimated from the complex band structure of BaTiO$_3$ in Fig. 5b. For simplicity we use in our estimate the average positions of the Fermi level for the two polarization states, indicated by the vertical lines in Fig. 5b. Taking the corresponding values of $\kappa$ we find that the exponential factor $\exp[-2\kappa d]$ changes by a factor of 2.2 for the $\Delta_1$ states and by a factor of 2.5 for the $\Delta_5$ states as the polarization changes from right to left, indicating that the dominant source of TER in these junctions is the change in complex band structure induced by the left-right asymmetry of the ferroelectric displacements in the barrier.

Our results clearly demonstrate the strong effect of ferroelectricity on electron and spin transport properties of multiferroic tunnel junctions comprising ferromagnetic electrodes and a ferroelectric barrier. Using SrRuO$_3$/BaTiO$_3$/SrRuO$_3$ junctions with asymmetric interfaces as a model system we have predicted the co-existence of tunnelling magnetoresistance and tunnelling electroresistance effects, indicating that magnetic tunnel junctions with ferroelectric barriers may serve as four-state resistance devices. These results reveal the exciting prospects of such multiferroic tunnel junctions for application in multilevel non-volatile memories, tuneable electric and magnetic field sensors, and multifunctional resistive switches. We hope that these results will further stimulate experimental efforts in studying magnetic tunnel junctions with ferroelectric barriers.

**Acknowledgements** The authors thank Chang-Beom Eom, Mark Rzchowski, Stefan Blügel, and Daniel Wortmann for helpful discussions, and Andrea Dal Corso for his assistance with the conductance code. This work was supported by NRI, NSF MRSEC (grant No. DMR-0820521), and ONR (grant No. N00014-07-1-1028). Work at UPR was supported by IFN (NSF grant No. 0701525). Work at ECNU was sponsored by NSFC (grant No. 50832003). Work at SISSA was sponsored by PRIN (Cofin 2006022847). Computations were performed at the Research Computing Facility (UNL) and the Center for Nanophase Materials Sciences (ORNL).

[1] Chappert, C., Fert, A. & Van Dau, F.N. *Nature Mater.* **6**, 813-823 (2007).
[2] Baibich, M.N. *et al. Phys. Rev. Lett.* **61**, 2472-2475 (1988).
[3] Binash, G., Grünberg, P., Saurenbach, F. & Zinn, W. *Phys. Rev. B* **39**, 4828-4830 (1989).
[4] Moodera, J.S., Kinder, L.R., Wong, T.M. & Meservey, R. *Phys. Rev. Lett.* **74**, 3273-3276 (1995).
[5] Parkin, S. S. P., Kaiser, C. Panchula, A. Rice, P. M. & Hughes B. *Nature Mater.* **3**, 862-867 (2004).
[6] Yuasa, S., Nagahama, T., Fukushima, A., Suzuki, Y. & Ando, K. *Nature Mater.* **3**, 868-871 (2004).
[7] Tsymbal, E.Y. & Pettifor, D.G. *Solid State Physics*, eds. H. Ehrenreich and F. Spaepen, Vol. **56** (Academic Press, 2001) pp.113-237.
[8] Tsymbal, E.Y., Mryasov, O.N. & LeClair, P.R. *J. Phys.: Condensed Matter* **15**, R109-R142 (2003).
[9] Eerenstein, W., Mathur, N.D. & Scott, J.F. *Nature* **442**, 759-765 (2006).
[10] Ramesh, R. & Spaldin, N.A. *Nature Mater.* **6**, 21-29 (2007).
[11] Bibes M. & Barthélémy A. *Nature Mater.* **7**, 425-426 (2008).
[12] Tsymbal, E.Y. & Kohlstedt, H. *Science* **313**, 181-183 (2006).
[13] Ahn, C. H., Rabe, K. M. & Triscone, J.-M. *Science* **303,** 488-491 (2004).
[14] Dawber, M., Rabe, K.M. & J.F. Scott *Rev. Mod. Phys.* **77**, 1083-1130 (2005).
[15] Lichtensteiger, C., Dawber, M. & Triscone, J.-M. *Physics of ferroelectrics: A modern perspective*, Series: Topics in Applied Physics , Vol. **105**, eds. Rabe, K.M., Ahn, C. H. & Triscone, J.-M. (Springer, 2007), pp.305-336.

16. Bune, A.V. *et al*. *Nature* **391**, 874-877 (1998).
17. Tybell, T., Ahn, C.H. & Triscone, J.-M. *Appl. Phys. Lett.* **75**, 856-858 (1999).
18. Fong, D.D. *et al*. *Science* **304**, 1650-1653 (2004).
19. Lichtensteiger, C., Triscone, J.M., Junquera, J. & Ghosez, Ph. *Phys. Rev. Lett.* **94**, 047603 (2005).
20. Tenne, D.A. *et al*. *Science* **313**, 1614-1616 (2006).
21. Jia, C.L. *et al.*, *Nature Mater.* **6**, 64-69 (2007).
22. Junquera, J. & Ghosez, Ph. *Nature* **422**, 506-509 (2003).
23. Sai, N., Kolpak, A.M. & Rappe, A.M. *Phys. Rev. B* **72**, 020101 (2005).
24. Duan, C.-G., Sabiryanov, R.F., Mei, W.N., Jaswal, S.S. & Tsymbal, E.Y. *Nano Lett.* **6**, 483 (2006).
25. Gerra, G., Tagantsev, A.K., Setter, N. & Parlinski, K. *Phys. Rev. Lett.* **96**, 107603 (2006).
26. Duan, C.-G. Jaswal, S.S. & Tsymbal, E.Y. *Phys. Rev. Lett.* **97**, 047201 (2006).
27. Velev, J.P., Dowben, P.A., Tsymbal, E.Y., Jenkins, & Caruso, A. *Surface Science Reports* **63**, 400-425 (2008).
28. Esaki, L., Laibowitz, R. B. & Stiles, P. J. *IBM Tech. Discl. Bull.* **13**, 2161 (1971).
29. Rodríguez Contreras, J., Kohlstedt, H., Poppe, U., Waser, R., Buchal, C. & Pertsev, N. A. *Appl. Phys. Lett.* **83**, 4595 (2003).
30. Zhuravlev, M.Y., Sabirianov R.F., Jaswal, S.S. & Tsymbal, E.Y. *Phys. Rev. Lett.* **94**, 246802 (2005).
31. Velev, J.P., Duan, C.-G., Belashchenko, K.D., Jaswal, S.S. & Tsymbal, E.Y. *Phys. Rev. Lett.* **98**, 137201 (2007).
32. Zhuravlev, M.Y., Jaswal, S.S., Tsymbal, E.Y. & Sabirianov, R. F. *Appl. Phys. Lett.* **87**, 222114 (2005).
33. Velev, J.P., Duan, C.-G., Belashchenko, K.D., Jaswal, S.S. & Tsymbal, E.Y. *J. Appl. Phys.* **103**, 07A701 (2008).
34. Gajek, M., Bibes, M., Fusil, S., Bouzehouane, K., Fontcuberta, J., Barthélémy, A. & Fert. A. *Nat. Mater.* **6**, 296 (2007).
35. Hill, N.A. *J. Phys. Chem. B* **104**, 6694-6709 (2000).
36. Petraru, A. et al. *Appl. Phys. Lett.* **93**, 072902 (2008).
37. Eom C.B. & Rzchowski, M.S., unpublished.
38. Giannozzi, P. *et al*. http://www.quantum-espresso.org.
39. Kresse, G. & Furthmüller, J. *Phys. Rev. B* **54** 11169 (1996); VASP (http://cms.mpi.univie.ac.at/vasp/).
40. The exchange-correlation potential is treated in the Perdew-Burke-Ernzerhof (PBE) generalized gradient approximation. The energy cut-off of 500 eV is used for the plane wave expansion and a 10x10x1 Monkhorst Pack grid for k-point sampling. The convergences over both cut-off energy and k-point sampling have been tested. Structural relaxations are performed until the Hellman-Feynman forces on atoms become less than 10 meV/Å.
41. The electrostatic potential profile is obtained as the solution of the Poisson equation for the self-consistent charge density and then averaged over the unit cell.
42. Gerra, G., Tagantsev, A.K. & Setter, N. *Phys. Rev. Lett.* **98**, 207601 (2007).
43. Choi, H.J. & Ihm, J. *Phys. Rev. B* **59**, 2267 (1999).
44. Smogunov, A., Dal Corso, A., & Tosatti, E. *Phys. Rev. B* **70**, 045417 (2004).
45. Three unit cells of $SrRuO_3$ on each side of the barrier were found to be sufficient to reproduce a bulk-like potential on both sides of the scattering region. Transmission and reflection matrices are then obtained by matching the wave functions in the scattering region to appropriate linear combinations of the Bloch states in the left and right leads. The zero-bias conductance is evaluated by integrating the electron transmission for states at the Fermi level over the two-dimensional Brillouin zone using a uniform 100x100 $\mathbf{k}_\parallel$ mesh. Due to the well known problem of incomplete self-interaction cancellation in GGA, the band gap of BaTiO3 is smaller than the experimental value of 3.2 eV. An increased value of the band gap will change the calculated transmission quantitatively, but not qualitatively.
46. TMR ratio is limited to ±100%. Within the conventional definition, $\text{TMR} = (G_{\uparrow\uparrow} - G_{\uparrow\downarrow})/G_{\uparrow\downarrow}$, the values of magnetoresistance may be infinite.
47. Mavropoulos, P., Papanikolaou, N. & Dederichs, P.H. *Phys. Rev. Lett.* **85**, 1088-1091 (2000).
48. Belashchenko, K.D., Tsymbal, E.Y., van Schilfgaarde, M., Stewart, D., Oleinik, I.I. & Jaswal, S.S. *Phys. Rev. B* **69**, 174408 (2004).